\documentclass[a4paper,11pt]{article}
\usepackage[utf8]{inputenc}
\usepackage[T1]{fontenc}
\usepackage{graphicx}
\usepackage{a4wide}
\usepackage{newtxtext}
\usepackage{newtxmath}
\usepackage[font=footnotesize,labelfont=bf]{caption}
\usepackage{amsmath}
 
\usepackage{booktabs}
\usepackage{longtable}
\usepackage{float}
\usepackage{authblk}

\usepackage{xurl}
\usepackage[numbers,super,sort&compress]{natbib}
\usepackage[hidelinks]{hyperref}

\usepackage{verbatim}
\newcommand{\detailtexcount}[1]{%
  \immediate\write18{texcount -merge -sum -q main.tex > main.wcdetail }%
  \verbatiminput{main.wcdetail}%
}

\usepackage{lineno} 
\usepackage{verbatim}

\makeatletter
\newcommand*{\addFileDependency}[1]{
  \typeout{(#1)}%
  \@addtofilelist{#1}%
  \IfFileExists{#1}{}{\typeout{No file #1.}}%
}
\makeatother

\title{Sequence models reveal diagnosis accumulation pathways beyond comorbidity burden in population-scale hospital data}

\author[1,2,3]{Katharina Ledebur}
\author[1,2]{Mitja Devetak} 
\author[1,2,3,4]{Peter Klimek*} 
\date{}

\affil[1]{Complexity Science Hub Vienna, Metternichgasse 8, A-1030 Vienna, Austria}
\affil[2]{Supply Chain Intelligence Institute Austria, Metternichgasse 8, A-1030 Vienna, Austria}
\affil[3]{Institute of the Science of Complex Systems, CeDAS, Medical University of Vienna, Spitalgasse 23, A-1090 Vienna, Austria}
\affil[4]{Division of Insurance Medicine, Department of Clinical Neuroscience, Karolinska Institutet, Stockholm, Sweden}
\affil[*]{klimek@csh.ac.at}
\begin{document}
\maketitle
\begin{abstract}
Aging trajectories vary among individuals of similar age and disease burden. Comorbidity indices, e.g. the Elixhauser index, summarize conditions cross-sectionally, but discard the timing, sequence, and pace of morbidity accumulation. Here we ask whether longitudinal hospital diagnosis histories contain information beyond age, sex, and comorbidity burden, and where it is concentrated.

Using 13 years of Austrian inpatient data covering 7.4 million patients, we trained a visit-level contrastive transformer to encode diagnosis sequences and inter-admission timing into patient-history embeddings. In a downstream cohort of 1.7 million individuals, embeddings improved prediction over the Elixhauser-based comorbidity model for 93 of 131 incident ICD-10 disease-block outcomes, with a modest median AUC gain of 0.006. Gains concentrated in mental, musculoskeletal, nervous system, and metabolic disorders. We then evaluated event-free survival, defined as remaining alive without accumulating a second unrecorded ICD-10 disease block. The embedding model achieved an AUC of 0.726 versus 0.722 for the comorbidity model. However, among patients with similar age, sex, and comorbidity-model risk, those assigned high residual risk had 132--183 fewer event-free days over five years and observed event rates comparable to low-residual-risk patients more than a decade older.

Together, these findings link the embedding's signal to the breadth, recency, and pace of prior disease accumulation.

\end{abstract}

\section*{Introduction}
Even among individuals of a similar age and disease burden, people experience different future health trajectories~\cite{ferrucci_heterogeneity_2021, nguyen_health_2021, calderon-larranaga_health_2021}. Understanding why and how this occurs is a central question in aging research, and requires tools capable of capturing the dynamic and longitudinal nature of the aging process.

Sequence models trained on electronic health records (EHRs) now produce patient representations from longitudinal diagnosis histories at population scale~\cite{siebra_transformers_2024}. 
Several approaches using transformer-based models show that pretraining on large clinical corpora yields transferable patient representations and improves multi-disease risk prediction relative to models that do not leverage longitudinal pretraining~\cite{ li_behrt_2020, huang_clinicalbert_2020, rasmy_med-bert_2021, li_hi-behrt_2023, kraljevic_foresightgenerative_2024,yang_transformehr_2023,moen_towards_2025,alhumaidi_use_2025}. Newer population-scale models extend this idea to explicit modeling of disease trajectories and competing risks~\cite{savcisens_using_2024}. SurvivEHR introduces a time-to-event pretraining objective for multimorbidity forecasting, while Delphi-2M learns generative disease progressions and enables simulation of long-term future health states~\cite{gadd_survivehr_nodate, shmatko_learning_2024}. Related sequence models predict outcomes such as early cancer detection and biological aging~\cite{placido_deep_2023,wang_full_2025}.  

The clinical relevance of this new type of prediction models is not yet clear. From a methodological perspective, they are trained to carry out tasks such as next-token prediction and masked language modeling on diagnostic sequences. Such objectives capture aspects of future diagnosis occurrence, but they are rarely evaluated against trajectory-level outcomes that are central to studying aging-related health, such as remaining alive while delaying multimorbidity accumulation over a defined period of time. Furthermore, it is not clear whether transformer-based models offer improved accuracy over established statistical tools to assess such outcomes. 

The standard clinical tool for quantifying cumulative disease burden is the Elixhauser comorbidity index, a weighted sum of 31 condition categories derived from administrative claims~\cite{elixhauser_comorbidity_1998}. Originally developed to predict in-hospital mortality and resource use from single-admission discharge records, it has since become one of the most widely used diagnosis-based case-mix adjustment tools in outcomes research~\cite{sharabiani_systematic_2012, austin_why_2015, van_walraven_modification_2009}. However, comorbidity indices are fundamentally cross-sectional summaries. They capture which conditions are present at a given time, but not the timing, sequence, or pace with which disease accumulates. This limitation may be particularly important for studying aging, which is increasingly understood as a latent, dynamic, and multidimensional process, where longitudinal trajectories can reveal clinically meaningful heterogeneity beyond static disease counts~\cite{li_large_2025,calderon-larranaga_understanding_2025, cezard_studying_2021,newman_generating_2023}. 

We use a contrastive objective to learn a single patient-level embedding from longitudinal visit sequences. In this way, the representation is explicitly optimized for comparing longitudinal medical histories, rather than for predicting a specific supervised outcome. The resulting embeddings map discrete and heterogeneous medical histories into a shared vector space, so that patients with similar trajectories are placed closer together and all patients can be compared within the same representation space.

We ask how much, and in what way, longitudinal patient-history embeddings add information about future morbidity accumulation beyond age, sex, and established comorbidity burden. Using nationwide Austrian hospital claims data from 1997 to 2009, we train a visit-level transformer with contrastive self-supervision to encode ICD-10 diagnosis sequences and inter-admission timing into patient embeddings. We compare embedding-based models with models using only age and sex, and with models using age, sex, and the Elixhauser comorbidity score, across 131 incident ICD-10 disease-block outcomes and a composite event-free survival endpoint defined as remaining free of in-hospital death or a second previously unrecorded ICD-10 disease block during follow-up from 2010 to 2014.

The embeddings improve prediction for most of 131 incident ICD-10 disease-block outcomes, with most gains being incremental and larger gains concentrated in mental, musculoskeletal, nervous system, and metabolic disorders. To capture broader morbidity accumulation, we also evaluate time to a second previously unrecorded ICD-10 disease block or in-hospital death as a trajectory-level endpoint. For this endpoint, we analyze embedding residual risk (difference between predicted risk from the embedding and the comorbidity model) among patients with similar age, sex, and comorbidity-model predicted risk. Patients assigned higher residual risk by the embedding experience shorter event-free survival and have denser, broader, more recent, and more multisystem hospital histories, including more prior visits, greater diagnostic diversity, faster disease-block accumulation, and shorter time since last admission. These features describe how morbidity accumulates over time and are not represented in the Elixhauser score. Thus, longitudinal embeddings expose hidden morbidity-accumulation risk among patients who appear similar under static comorbidity summaries, even if there is little gain in predicting individual diagnoses.

\section*{Results}

Using a contrastive learning objective, we derived patient representations from sequences of hospital admissions before the 2010 landmark, spanning up to 13 years of prior history. The visit-level transformer encoded ICD-10 diagnoses, time since previous admission, and sex into a fixed-size patient embedding.

\begin{figure}
    \centering
    \includegraphics[width=0.99\linewidth]{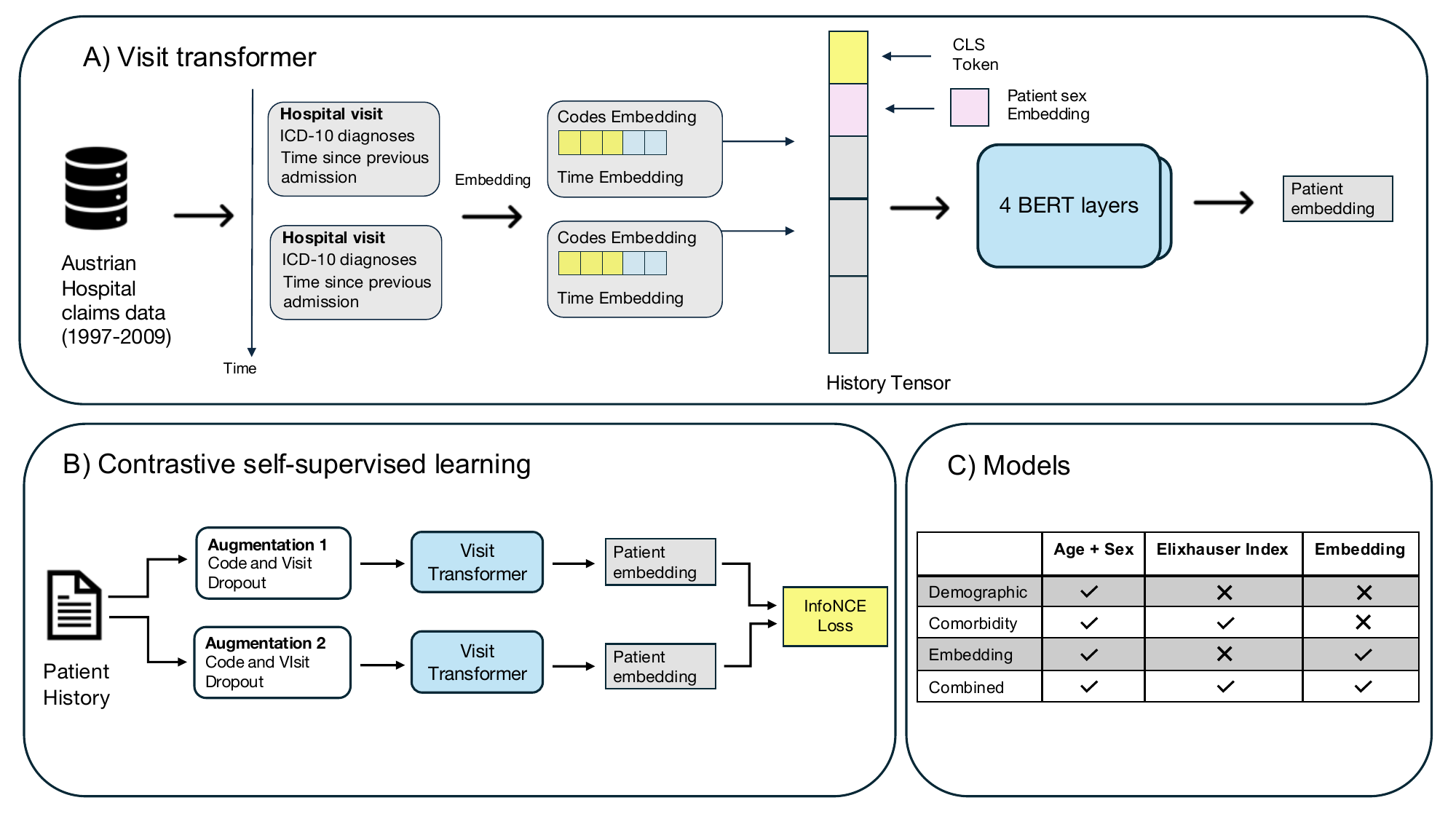}
    \caption{Overview of the visit transformer framework with contrastive self-supervised learning and downstream prediction models. (A) Visit-level transformer architecture applied to Austrian nationwide hospital claims data (1997-2009). Each hospital visit is represented by ICD-10 diagnosis embeddings and time since previous admission, processed through a four-layer BERT-style transformer to produce a patient-level embedding summarizing longitudinal history. (B) Contrastive self-supervised pretraining using dual stochastic augmentations of patient histories via code and visit dropout, optimized with an InfoNCE loss to learn robust patient representations without outcome labels. (C) Comparison of downstream prediction models for 131 five-year (2010--2014) incident ICD-10 disease-block outcomes from A00--N99, including demographic, comorbidity, embedding-based, and combined models. }
    \label{fig:visabs}
\end{figure}

At the 2010 landmark, we used these embeddings as inputs to downstream prediction models for future morbidity accumulation. We evaluated four model classes: demographic (age and sex), comorbidity (age, sex, and Elixhauser score), embedding (age, sex, and the learned patient-history embedding), and combined (all features). These models were benchmarked across incident ICD-10 disease-block outcomes and for a broader morbidity-accumulation endpoint defined as time to second previously unrecorded ICD-10 disease block or in-hospital death during follow-up (2010--2014) (Figure~\ref{fig:visabs}).

\subsection*{Longitudinal disease-history embeddings add predictive information beyond comorbidity burden}

\begin{figure}[H]
    \centering
    \includegraphics[width=0.95\linewidth]{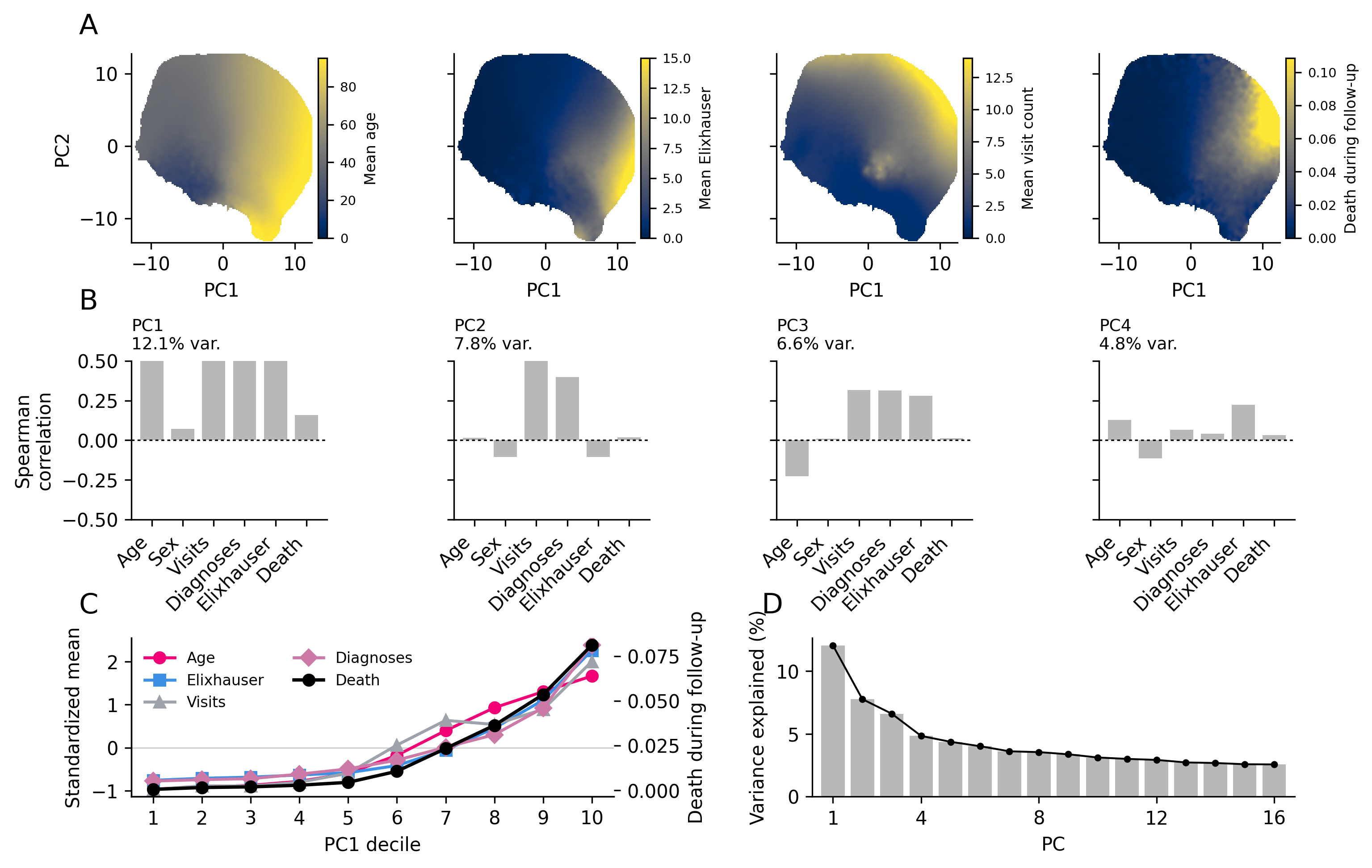}
    \caption{Clinical structure of the learned patient-history embedding. Patient-history embeddings were projected onto principal components (PCs) to characterize the structure learned by the self-supervised encoder. (A) Smoothed binned maps of the first two PCs show mean age at landmark (2010), mean Elixhauser comorbidity score, mean hospital visit count, and in-hospital death during follow-up across embedding space. Death is shown as the proportion of patients with recorded in-hospital  death during follow-up. (B) Spearman correlations between the first four PCs and selected clinical variables. PC signs were oriented so that the largest absolute displayed correlation was positive. (C) Clinical characteristics across deciles of PC1. Age at landmark, Elixhauser score, visit count, and diagnosis count are shown as standardized means; death during follow-up is shown as a proportion. (D) Variance explained by the first 16 principal components.}
    \label{fig:pc_analysis}
\end{figure}

Principal component analysis of the embeddings showed a dominant age, morbidity, and utilization gradient along PC1. However, the embedding did not collapse patient histories into a single scalar burden measure. Later components captured distinct mixtures of utilization, diagnosis diversity, comorbidity, and mortality, while Elixhauser score and visit count showed partly distinct gradients across the embedding space (Supplementary Fig.~\ref{fig:pc_analysis}). Thus, before any supervised outcome modeling, the representation had learned clinically interpretable structure that was not reducible to either cross-sectional comorbidity burden or healthcare utilization alone.

We tested whether the learned embeddings contained information about future disease risk beyond age, sex, and established comorbidity burden. Across 131 incident ICD-10 disease-block outcomes, the embedding model had higher discrimination than the demographic model for 101 outcomes, with a median AUC gain of 0.012; 87 outcomes had bootstrap confidence intervals for the AUC difference entirely above zero. Compared with the comorbidity model, the embedding model had higher discrimination for 93 outcomes, with a median AUC gain of 0.006; 70 outcomes had bootstrap confidence intervals entirely above zero.

Embedding gains over the comorbidity model varied across disease chapters. The largest median gains were observed for mental disorders (median AUC gain 0.030; 10 of 11 outcomes had higher AUC), musculoskeletal diseases (0.013; 6 of 6), nervous system diseases (0.012; 9 of 11), and metabolic diseases (0.011; 7 of 8). Gains were smaller for digestive, circulatory, respiratory, eye, and genitourinary diseases, and were small or absent for neoplasms, ear diseases, blood and immune disorders, infectious diseases, and skin diseases.

\begin{figure}[H]
    \centering
    \includegraphics[width=0.9\linewidth]{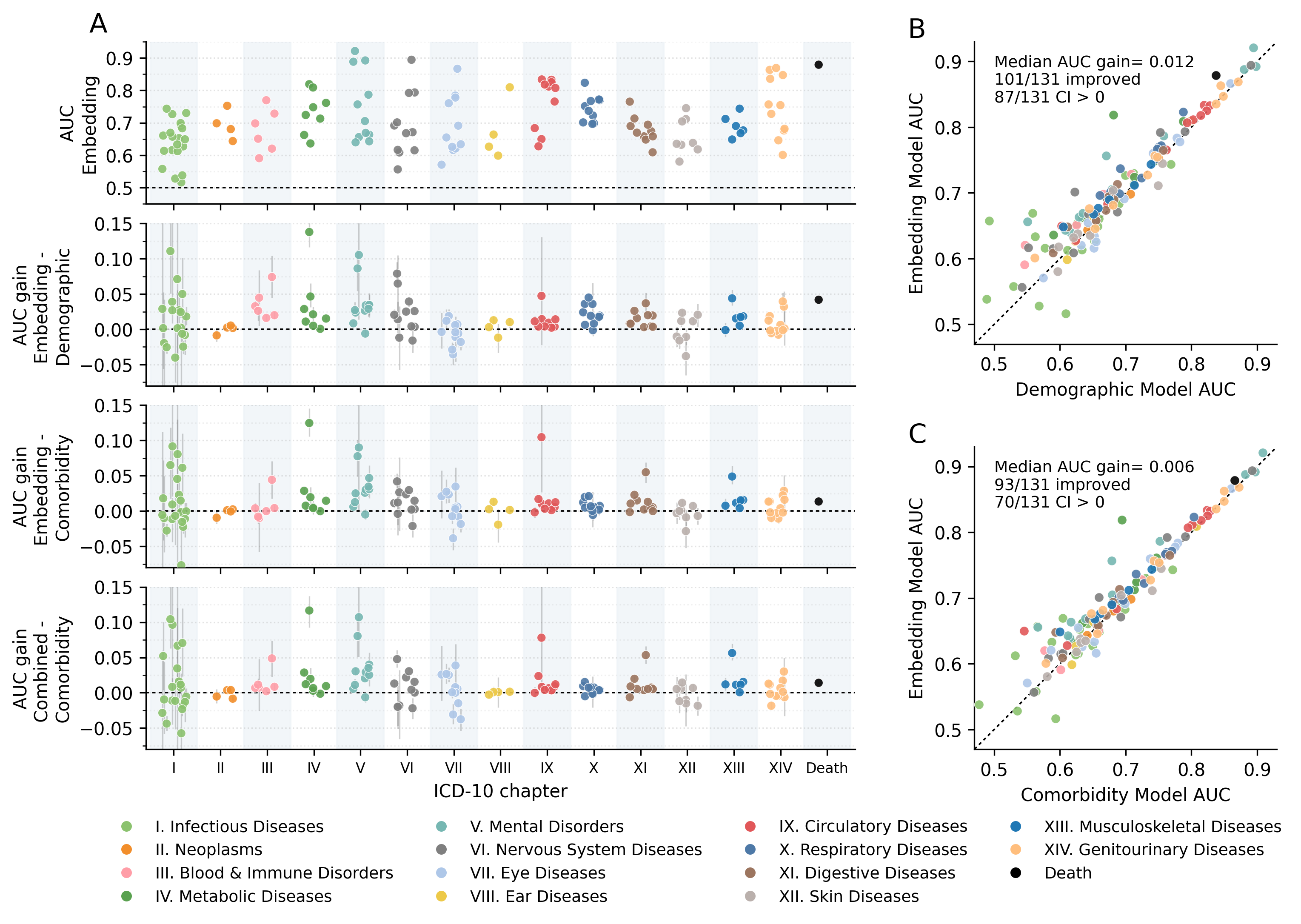}
\caption{Prediction of incident disease blocks from learned patient-history embeddings. Each point represents one incident ICD-10 disease-block outcome among patients free of the respective block at the 2010 landmark. The demographic model included age and sex; the comorbidity model included age, sex, and Elixhauser score; the embedding model included age, sex, and the learned patient-history embedding; and the combined model included age, sex, Elixhauser score, and the embedding. (A) Outcome-level AUCs and AUC differences by ICD-10 chapter. Points show individual outcomes, grouped by chapter; vertical lines show bootstrap 95\% confidence intervals for AUC differences. (B) AUC of the embedding model compared with the demographic model. (C) AUC of the embedding model compared with the comorbidity model. Dashed diagonals indicate equal AUC; points above the diagonal indicate higher AUC for the embedding model.}

    \label{fig:auc}
\end{figure}

\subsection*{Residual embedding risk identifies divergent trajectories}

We then evaluated whether embedding-based predictions stratified broader morbidity accumulation over follow-up. The primary event-free survival endpoint was remaining alive without accumulating a second previously unrecorded ICD-10 disease block. For this endpoint, the embedding model achieved an AUC of 0.726 compared with 0.722 for the comorbidity model, corresponding to an AUC gain of 0.0037 (95\% CI 0.0031--0.0042). The combined model showed a slightly larger improvement over the comorbidity model (AUC gain 0.0048, 95\% CI 0.0042--0.0053; Supplementary Table~\ref{tab:survival_endpoint_auc}).

We next asked whether the embedding model distinguished patients with similar demographic and Elixhauser-based risk but different future morbidity accumulation. For each held-out individual, we defined embedding residual risk as
\[
r_i = \hat p_i^{\mathrm{embedding}} - \hat p_i^{\mathrm{comorbidity}},
\]
where \(\hat p_i^{\mathrm{embedding}}\) is the predicted risk from the embedding model and \(\hat p_i^{\mathrm{comorbidity}}\) is the predicted risk from the model using age, sex, and Elixhauser score. Positive residual risk therefore indicates that the embedding model assigned higher risk than expected from the comorbidity model and vice versa.

\begin{figure}[H]
    \centering
    \includegraphics[width=0.9\linewidth]{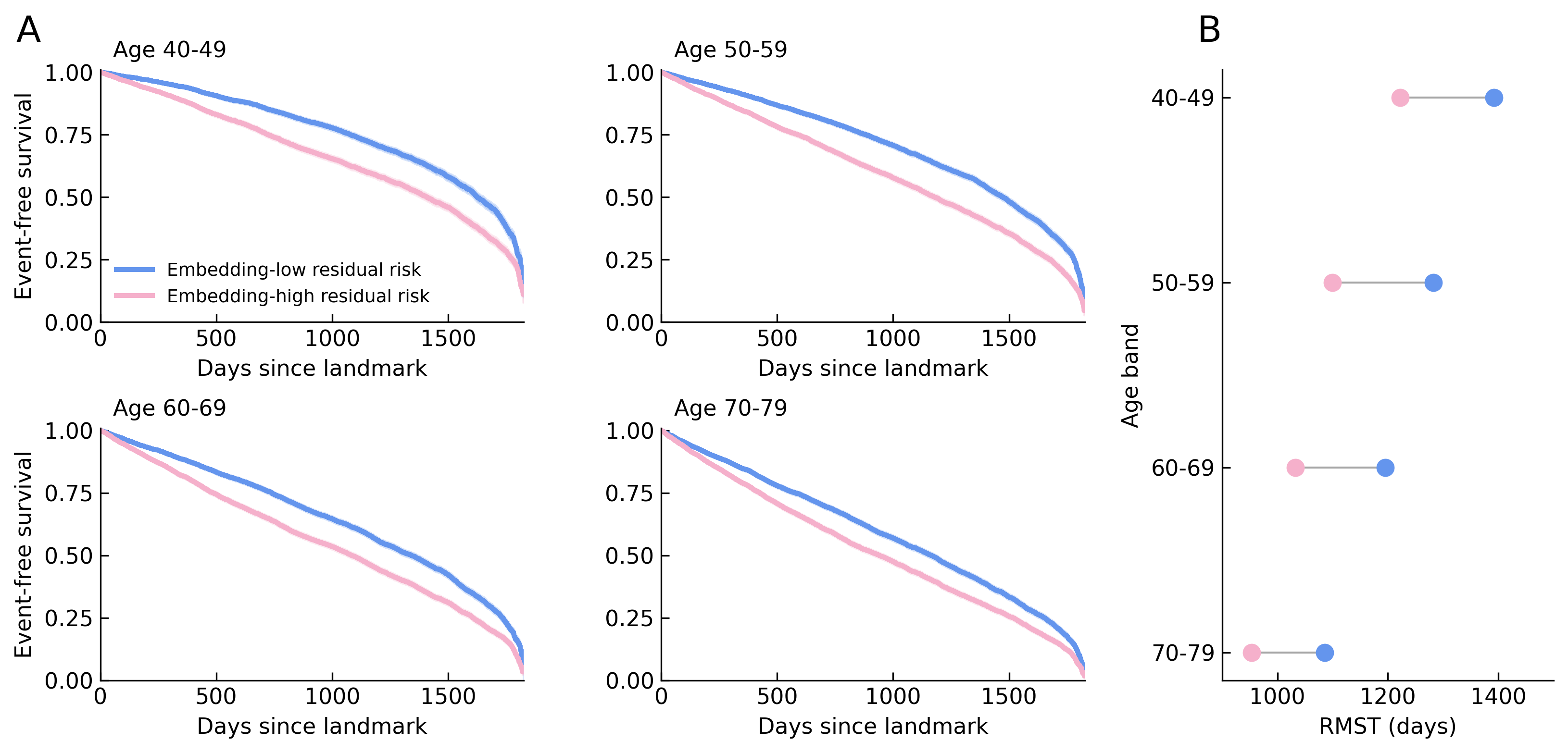}
    \caption{Embedding residual risk separates future event-free trajectories among patients with similar comorbidity-model risk. Embedding residual risk was defined as the difference between embedding-model and comorbidity-model predicted risk for second incident ICD-10 disease block or death. Residual-risk quintiles were assigned within strata of age band, sex, and comorbidity-model predicted-risk decile in the held-out test set. (A) Kaplan--Meier curves show time to second incident ICD-10 disease block or death for patients in the lowest and highest residual-risk quintiles within each age band. Blue indicates embedding-low residual risk and pink indicates embedding-high residual risk. (B) Restricted mean event-free survival time (RMST) over five years for embedding-low and embedding-high residual-risk groups.}

    \label{fig:kaplan}
\end{figure}

To compare patients with similar baseline risk, we assigned residual-risk quintiles within strata defined by age band, sex, and deciles of comorbidity-model predicted risk. We refer to patients in the lowest residual-risk quintile as embedding-low and patients in the highest residual-risk quintile as embedding-high. Kaplan--Meier curves comparing these two groups showed consistently lower event-free survival in embedding-high individuals across age bands (Figure~\ref{fig:kaplan}A). Over five years, embedding-high individuals had shorter restricted mean event-free survival than embedding-low individuals by 169.9, 183.0, 163.2, and 132.0 days at ages 40--49, 50--59, 60--69, and 70--79, respectively (Figure~\ref{fig:kaplan}B).

\begin{figure}[H]
    \centering
    \includegraphics[width=0.9\linewidth]{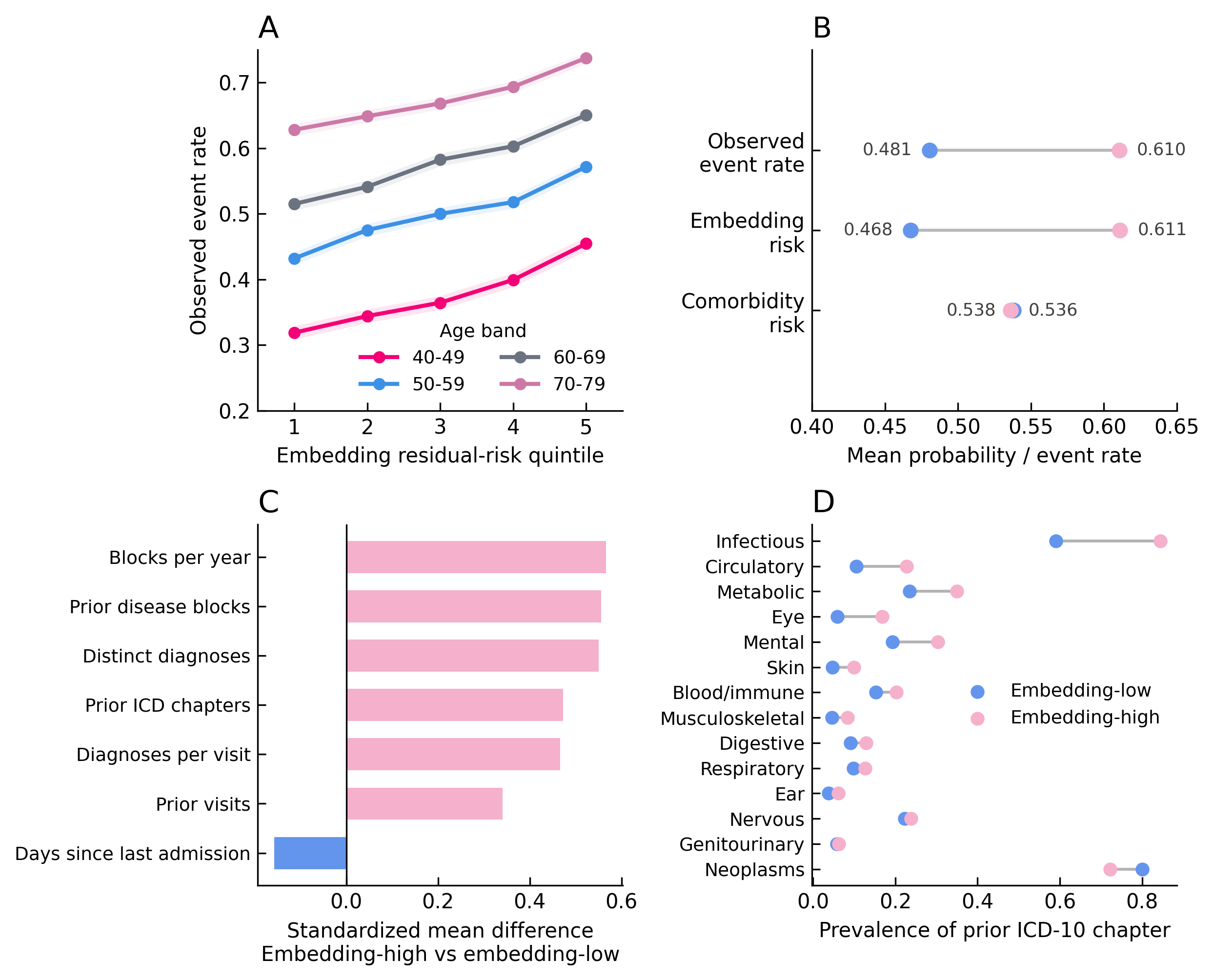}
    \caption{Embedding residual risk identifies heterogeneous future morbidity among patients with similar comorbidity-model risk. Embedding residual risk was defined as the difference between embedding-model and comorbidity-model predicted risk for second incident ICD-10 disease block or death. Residual-risk quintiles were assigned within age, sex, and comorbidity-model risk strata in the held-out test set. (A) Observed event rate across residual-risk quintiles by age band. (B) Mean comorbidity-model predicted risk, embedding-model predicted risk, and observed event rate in embedding-low and embedding-high individuals, defined as the lowest and highest residual-risk quintiles. (C) Standardized mean differences in pre-landmark history features comparing embedding-high with embedding-low individuals. Positive values indicate higher values in embedding-high individuals. (D) Prevalence of prior ICD-10 chapters recorded before the landmark in embedding-low and embedding-high individuals.
}

    \label{fig:residualrisk}
\end{figure}

Consistent with this time-to-event separation, observed event rates increased monotonically across residual-risk quintiles in all age bands (Figure~\ref{fig:residualrisk}A), indicating that the embedding captured risk-relevant variation even within groups of similar age, sex, and comorbidity-model predicted risk. Because these strata were defined using the comorbidity model, patients compared across residual-risk quintiles had similar predicted risk under the age-, sex-, and Elixhauser-based model. The two extreme groups had nearly identical mean comorbidity-model predicted risk (0.530 in embedding-high and 0.531 in embedding-low), but differed substantially in embedding-model predicted risk (0.600 versus 0.464) and observed event rate (60.6\% versus 48.6\%; Figure~\ref{fig:residualrisk}B).

Put on an age scale, embedding-high patients had event rates similar to embedding-low patients more than a decade older. For example, embedding-high patients aged 50, 55, 60, and 65 years resembled embedding-low patients aged about 64, 69, 73, and 77 years, respectively. The median age-equivalent difference across the overlapping age range was 12.9 years.

The clinical histories of these groups differed systematically before the landmark. Compared with embedding-low individuals, embedding-high individuals had more prior hospital visits (mean 5.12 versus 3.21), more distinct diagnoses (8.20 versus 4.58), more prior disease blocks (6.29 versus 3.80), more affected ICD-10 chapters (3.72 versus 2.72), more diagnoses per visit (2.13 versus 1.60), and a higher rate of disease-block accumulation per year (3.22 versus 2.04). They also had a shorter time since last admission (859 versus 1,099 days; Figure~\ref{fig:residualrisk}C). At the chapter level, embedding-high individuals had higher prior prevalence of infectious diseases (85.7\% versus 52.1\%), metabolic diseases (34.0\% versus 22.3\%), mental disorders (30.8\% versus 17.0\%), circulatory diseases (21.2\% versus 11.2\%), eye diseases (15.2\% versus 6.3\%), skin diseases (9.3\% versus 4.6\%), and musculoskeletal diseases (7.5\% versus 4.6\%). In contrast, prior neoplasm diagnoses were slightly less prevalent in the embedding-high group (72.7\% versus 75.9\%; Figure~\ref{fig:residualrisk}D). Together, these findings suggest that the embedding's added signal corresponded to measurable differences in pre-landmark history that are not encoded in the Elixhauser score, including diagnostic diversity and the breadth and pace of disease-block accumulation.

Because embeddings can encode the recency of hospital contact, we repeated the composite-endpoint benchmark after adding days since last pre-landmark admission to both the embedding and comorbidity models. Recency modestly improved the comorbidity model (AUC 0.724), but the embedding model retained a similar advantage over the recency-adjusted comorbidity model (AUC 0.729 versus 0.724; AUC gain = 0.0054, 95\% CI 0.0049--0.0059).

\section*{Discussion}

We show that longitudinal hospital diagnosis histories reveal heterogeneity in future morbidity accumulation that is not captured by age, sex, or Elixhauser comorbidity burden. The added signal was not a large uniform gain in discrimination. Instead, it appeared as outcome-specific improvements and, more importantly, as separation of future event-free trajectories among patients who appeared similar under a conventional comorbidity model. In residual-risk analyses, embedding-high individuals experienced 132--183 fewer event-free days over five years across age bands despite nearly identical comorbidity-model risk, and had observed event rates comparable to embedding-low individuals more than a decade older.

Across 131 incident ICD-10 disease-block outcomes, the embedding model had higher AUC than the demographic model for 101 outcomes and higher AUC than the Elixhauser-based comorbidity model for 93 outcomes, with median AUC gains of 0.012 and 0.006, respectively. Bootstrap confidence intervals for the AUC difference were entirely above zero for 87 outcomes in the comparison with the demographic model and for 70 outcomes in the comparison with the comorbidity model. Gains were concentrated in specific parts of the morbidity spectrum, especially mental disorders, musculoskeletal diseases, nervous system diseases, and metabolic diseases, and were small or absent for neoplasms, ear diseases, blood and immune disorders, infectious diseases, and skin diseases.

The modest size of the average gains should be interpreted in light of the strength of the baseline. Age, sex, and Elixhauser score already capture substantial prognostic information in administrative hospital data, and comorbidity indices have been extensively validated for mortality and resource-use prediction~\cite{austin_why_2015, van_walraven_modification_2009}. Their limitation is not that they are uninformative, but that they compress history. Diseases are represented as cross-sectional indicators with fixed weights, while temporal order, recurrence, density, and pace of accumulation are discarded. The broader morbidity-accumulation endpoint made this distinction visible. For second incident ICD-10 disease block or in-hospital death, the embedding model improved AUC over the comorbidity model from 0.722 to 0.726 (AUC gain = 0.0037, 95\% CI 0.0031--0.0042), but the residual-risk analysis revealed larger differences in event-free survival among patients with similar conventional risk. Thus, the value of the embedding lies less in uniformly increasing average discrimination than in identifying trajectory differences that are compressed by static comorbidity summaries.

The residual-risk analysis links the embedding signal to concrete differences in pre-landmark hospital histories. Because residual-risk groups were formed within strata of age, sex, and comorbidity-model predicted risk, embedding-high and embedding-low patients had nearly identical mean comorbidity-model risk (0.530 versus 0.531). Nevertheless, their observed event rates differed substantially (60.6\% versus 48.6\%), and their subsequent event-free trajectories diverged. The embedding was also not reducible to hospital utilization. In the unsupervised embedding analysis, visit count and Elixhauser score showed only partly overlapping gradients, indicating that the representation captured variation beyond how often patients were hospitalized and how much comorbidity had already been recorded.

Embedding-high patients had broader, denser, and more recent pre-landmark histories, with more prior hospital visits, more distinct diagnoses, more prior disease blocks, more affected ICD-10 chapters, faster disease-block accumulation, and shorter time since last admission. Their histories also differed in clinical composition, with higher prior prevalence of infectious, metabolic, mental, circulatory, eye, skin, and musculoskeletal chapters, and slightly lower prevalence of neoplasms. Together, these patterns suggest that the embedding captured a mode of morbidity accumulation defined by recency, density, and multisystem breadth rather than by any single diagnosis category.

The use of hospital-diagnosis trajectories alone also shapes how the magnitude of the results should be interpreted. This design enables direct comparison to the Elixhauser score, which is also diagnosis-based, but it does not measure the performance achievable with richer data sources such as outpatient diagnoses, medications, laboratory values, socioeconomic information, or lifestyle factors. We therefore cannot fully resolve whether the modest gains reflect limitations of sequence models for these outcomes or the limited clinical view provided by inpatient diagnoses. At the same time, recent population-scale sequence models using richer clinical and socioeconomic information have also reported incremental improvements over age- and sex-based baselines~\cite{savcisens_using_2024,shmatko_learning_2024}. This suggests that modest average gains may be typical when predicting broad morbidity outcomes from already informative baseline features. The central question is therefore not only whether sequence models increase average AUC, but whether they reveal clinically meaningful heterogeneity that static burden summaries obscure.

Several limitations should be noted. First, the analysis is based on hospital claims data, which capture diagnosed and treated inpatient conditions but miss primary care, outpatient diagnoses, medication use, laboratory measurements, lifestyle factors, and socioeconomic circumstances. The embedding therefore reflects the subset of health history visible through hospital contact, and conditions managed entirely outside inpatient care will be underrepresented. Nationwide coverage and a long observation window partially offset this limitation by capturing serious or recurring conditions across the full population.

Second, diagnosis-based outcomes depend on healthcare use and coding practices. Patients with more frequent or more recent hospital contact may have more opportunities for new diagnoses to be recorded, which can influence incident disease-block outcomes. This is particularly relevant because embedding-high individuals had shorter time since last admission in the residual-risk analysis. However, in a sensitivity analysis for the composite endpoint, adding days since last pre-landmark admission to both the embedding and comorbidity models did not materially change the embedding advantage (AUC 0.729 versus 0.724; AUC gain = 0.0054, 95\% CI 0.0049--0.0059). This suggests that the signal captured by the embedding was not explained solely by hospital-contact recency, although residual surveillance effects may remain.

Third, the downstream prediction cohort required at least one hospital encounter during follow-up and no recorded admission before 2000. This design ensured observable pre-landmark history and follow-up contact, but it also restricts inference to individuals visible in the inpatient system during the study window. Fourth, in-hospital death was observed, whereas deaths outside hospital were not captured. Fifth, the analysis was conducted within a single national health system and requires external validation before clinical use.

Finally, although prediction performance was evaluated on held-out data, the analyses used to interpret what the embedding captured were descriptive. The residual-risk characterization, prior-diagnosis prevalence comparisons, and principal-component analyses identify associations between diagnosis patterns and future morbidity accumulation in this population, but they do not establish whether those associations reflect modifiable disease processes, healthcare-contact patterns, or fixed patient characteristics.

In summary, longitudinal hospital diagnosis histories contain information about future morbidity accumulation that is not fully captured by age, sex, or Elixhauser comorbidity burden. Contrastive sequence models can extract this information at population scale and identify patients whose future trajectories differ substantially despite similar apparent burden. For aging research, such representations may help move beyond static counts of disease toward trajectory-based descriptions of morbidity accumulation, multimorbidity progression, and healthspan heterogeneity.


\section*{Methods}
\subsection*{Data and Cohort}

We used nationwide Austrian inpatient claims data to construct longitudinal hospital histories from 1997 to 2009. For each patient, we represented the history as a chronological sequence of hospital visits, including sex, time since the previous visit, and all primary and secondary ICD-10 diagnoses recorded at each visit. In-hospital death was encoded as an additional diagnosis code when it occurred. The self-supervised representation model was trained on 7,418,519 patients.

Downstream prediction analyses used 2010 as the landmark year and evaluated outcomes during follow-up from 2010 to 2014. The downstream prediction cohort comprised 1,734,479 individuals with complete feature availability across age, sex, Elixhauser score, and patient-history embeddings, at least one hospital encounter during follow-up, and no recorded hospital admission before 2000. Outcome-specific at-risk cohorts and event definitions are described below.

\subsection*{Patient History Encoder.}

\subsubsection*{Notation}

Each patient $i \in \{1,\dots,N\}$ has a history given by a chronological sequence of visits. Each visit contains a set of primary and secondary diagnoses encoded as ICD-10 codes, augmented with a dedicated code for death, and the time elapsed since the previous visit. Let $\mathcal{V}$ denote the vocabulary of all possible codes and let $|\mathcal{V}|$ be its size. Codes are represented by integer indices in $\{0,\dots,|\mathcal{V}|-1\}$, where one index is reserved for padding.

We represent each history as a fixed-size tensor by capping the number of visits at $128$ and the number of codes per visit at $64$. If a patient has fewer visits or a visit has fewer codes, the remaining entries are filled with the padding code and a padding time delta. In the data, $0.015\%$ of patients have more than $128$ visits and no visit contains more than $64$ diagnosis codes, so this truncation discards little information.

Sex is encoded as $s_i \in \{0,1\}$ with $0$ for male and $1$ for female. The visit-code matrix is
\[
X_i \in \{0,\dots,|\mathcal{V}|-1\}^{128 \times 64},
\]
including the padding index. Per-visit time deltas are
\[
\Delta_i \in \{0,\dots,30\}^{128},
\]
measured in months since the previous visit and capped at $30$, with the first visit set to $0$. The patient representation is $(X_i,\Delta_i,s_i)$. The goal of pretraining is to learn a low-dimensional embedding of the medical history.

\subsubsection*{Visit token construction}

We map each medical code ID to a learned embedding in $\mathbb{R}^{8}$ and set the embedding of the padding code to the zero vector by construction. We then project code embeddings to the transformer width which is $32$ using a single linear layer,

\begin{equation}
\tilde e(x) = W_c e(x) + b_c,\qquad W_c\in\mathbb{R}^{32\times 8},\; b_c\in\mathbb{R}^{32}.
\end{equation}

Empirically, this two-step embedding parameterization yielded more stable training than a direct 32-dimensional lookup. Within each visit, we aggregate the (projected) code embeddings into a single visit token using a masked mean that ignores padding codes. Concretely, for visit $t$ we average only over the non-padding codes present in that visit, and if a visit is empty after padding we map it to the zero vector.

Time is incorporated by embedding the binned time delta for each visit into $\mathbb{R}^{32}$ and adding it to the visit token. Again padding deltas map to the zero vector and hence don't contribute any information. The result is one token per visit that carries both a pooled code signal and a coarse elapsed-time signal.

This creates a per-visit embedding token which is a vector of dimension $32$.

\subsubsection*{Sequence assembly and masking}

The transformer input sequence starts with two special tokens: a learned CLS token and a learned sex token obtained from a two-entry lookup table indexed by $s_i$. These are followed by the $128$ visit tokens. An attention mask restricts self-attention to the prefix that corresponds to the true number of visits in the patient history, so padded visit positions do not participate.

\subsubsection*{BERT encoder and patient embedding}

We use a BERT-style encoder-only transformer with $4$ layers, $4$ attention heads, hidden size $32$, and feed-forward width $256$, with residual connections, dropout, and layer normalization~\cite{devlin_bert_2019}. We denote this encoder as $\mathrm{BERT}_{\theta}(\cdot)$ with parameters $\theta$. Given the masked input sequence $X_i^{\mathrm{seq}}$ and mask $M_i$, the encoder returns a sequence of contextualized token states
\[
Y_i = \mathrm{BERT}_{\theta}(X_i^{\mathrm{seq}}, M_i), 
\qquad 
Y_i \in \mathbb{R}^{(2+128)\times 32}.
\]

We define the patient embedding as the sum of the final CLS state and the final sex-token state. If $Y_i^{\mathrm{CLS}} \in \mathbb{R}^{32}$ and $Y_i^{\mathrm{SEX}} \in \mathbb{R}^{32}$ denote these two output states, the final embedding is
\[
Z_i = Y_i^{\mathrm{CLS}} + Y_i^{\mathrm{SEX}}, 
\qquad 
Z_i \in \mathbb{R}^{32}.
\]
No additional projection head is applied, so the contrastive representation used in the loss is exactly $Z_i$. This keeps the representation dimension at $32$ and avoids parameters that are only used during pretraining.

\subsubsection*{Two-view augmentation}

Training is self-supervised from two stochastic views of each patient. Each view is formed by subsampling visits, applying code dropout within the retained visits, and packing the retained visits to the front so that the valid prefix is contiguous. This prevents padded visits from carrying information.

Visit subsampling removes a random subset of visits from the patient history (without replacement). For view $r \in \{1,2\}$, a keep ratio $\rho^{(r)}$ is drawn uniformly from an epoch-dependent interval $[\rho_{\min}(e),1]$. If patient $i$ has $\ell_i$ valid visits, the number of retained visits in that view is
\[
\tilde{\ell}^{(r)}_i = \max\!\Bigl(k_{\min}(e),\ \mathrm{round}\bigl(\rho^{(r)} \ell_i\bigr)\Bigr).
\]
During epochs $1$ to $10$, $\rho_{\min}(e)=0.7$ and $k_{\min}(e)=3$. During epochs $11$ to $30$, $\rho_{\min}(e)$ decreases linearly from $0.7$ to $0.5$ and $k_{\min}(e)=2$.

Within each retained visit, code dropout is applied. For a retained visit with $K$ non-padding codes, a dropout fraction $\delta^{(r)}$ is drawn uniformly from an epoch-dependent interval $[\delta_{\min}(e),\delta_{\max}(e)]$, and
\[
D = \mathrm{round}\bigl(\delta^{(r)} K\bigr)
\]
of the $K$ codes are replaced by the padding code. When $K>1$, we enforce $D \le K-1$ so that at least one non-padding code remains. During epochs $1$ to $10$, $\delta^{(r)} \sim \mathrm{Unif}(0,0.05)$. During epochs $11$ to $30$, the interval endpoints are shifted linearly until $\delta^{(r)} \sim \mathrm{Unif}(0.10,0.20)$. The two views therefore correspond to partially observed and differently masked realizations of the same history.

After masking, the retained visits are packed into the leading $128$ visit slots in chronological order. The valid sequence is then a contiguous prefix of length $2+\tilde{\ell}^{(r)}_i$ (CLS, sex, and retained visits), so the attention mask is a single contiguous block and padded positions do not appear in the middle of the sequence.

\subsubsection*{Contrastive objective}

Let $Z_i^{(1)}$ and $Z_i^{(2)}$ be the $32$-dimensional patient embeddings obtained from the two views of patient $i$ in a batch of size $B$. We form $\mathbf{z}_i^{(1)}$ and $\mathbf{z}_i^{(2)}$ by $\ell_2$ normalizing these embeddings, $\mathbf{z}_i^{(r)} = Z_i^{(r)} / \|Z_i^{(r)}\|_2$. The training objective is the symmetric InfoNCE loss
\[
\mathcal{L}_{\text{total}} = -\frac{1}{2B} \sum_{i=1}^{B} \left[ 
\log \frac{\exp(\mathbf{z}_i^{(1)} \cdot \mathbf{z}_i^{(2)} / \tau)}{\sum_{j=1}^{B} \exp(\mathbf{z}_i^{(1)} \cdot \mathbf{z}_j^{(2)} / \tau)}
+ 
\log \frac{\exp(\mathbf{z}_i^{(2)} \cdot \mathbf{z}_i^{(1)} / \tau)}{\sum_{j=1}^{B} \exp(\mathbf{z}_i^{(2)} \cdot \mathbf{z}_j^{(1)} / \tau)}
\right].
\]
For each patient, the embedding from one view is trained to match the embedding from the other view of the same patient, while embeddings of other patients in the batch act as negatives. Since embeddings are $\ell_2$ normalized, dot products correspond to cosine similarities. We use a fixed temperature $\tau=0.07$.

\subsubsection*{Optimization, scheduling, and early stopping}

Training runs for 30 epochs on a NVIDIA L40 GPU. We optimize all parameters with Adam using a maximum learning rate of $10^{-2}$. We apply a cosine learning-rate schedule with linear warmup over an estimated total step budget $T$; the warmup is capped at 100 steps or one tenth of the total steps, whichever is smaller. Gradients are clipped by global norm to 1.0 to limit rare large updates and control the inherent instability of the InfoNCE loss. We use batches of 4096.

\subsubsection*{Embedding export}

After pretraining, we run the encoder once over the full dataset without shuffling to compute a 32-dimensional embedding $Z_i$ for each patient $i$. 

\subsection*{Prediction Benchmarks Across ICD-10 Disease Blocks}

We evaluated whether the learned patient-history embedding contained information about future morbidity beyond demographic variables and established comorbidity burden. Within the downstream prediction cohort, we defined an outcome-specific at-risk cohort for each ICD-10 disease block as individuals without a recorded diagnosis from that block before the 2010 landmark. The outcome was incident diagnosis of the block during follow-up. Death during follow-up was evaluated as an additional outcome.

The Elixhauser comorbidity score was used as the diagnosis-based clinical burden reference because it summarizes a broad set of comorbid conditions from administrative claims and has been extensively validated for inpatient mortality and resource-use prediction~\cite{sharabiani_systematic_2012, van_walraven_modification_2009, sharma_comparing_2021}. Scores were derived from ICD-10 codes recorded before the landmark using the Quan adaptation, as implemented in \texttt{comorbidipy} version 0.8.0~\cite{quan_coding_2005,noauthor_comorbidipy_2026}.

For each outcome, individuals were randomly split into training and test sets (80/20), stratified by outcome status. We trained separate multilayer perceptron classifiers using four feature sets: age and sex (demographic model), age, sex, and Elixhauser score (comorbidity model), age, sex, and the learned embedding (embedding model), and age, sex, Elixhauser score, and the learned embedding (combined model). All input features were standardized before model fitting. The MLP used hidden layers of size 64, 128, and 64 with ReLU activation. The classifier was trained with Adam and early stopping based on validation performance.

Model discrimination was evaluated on the held-out test set using the area under the receiver operating characteristic curve (AUC). $\Delta$ AUCs were calculated between paired models evaluated on the same test individuals. 

\subsection*{Morbidity-Accumulation and Event-Free Survival Analysis}

To assess whether the embeddings stratify broader aging-related morbidity accumulation, we defined event-free survival as remaining alive without accumulating a second previously unrecorded ICD-10 disease block during the 2010--2014 follow-up window. For each patient, we identified all disease blocks recorded before the 2010 landmark using primary and secondary diagnoses. During follow-up, a new disease block was defined as the first occurrence of a block not previously recorded before the landmark. The event time was the earlier of the second new disease-block date or in-hospital death. Patients without an event were censored at their last observed hospital contact or at the end of follow-up. In sensitivity analyses, we also evaluated death alone and a broader composite endpoint of first new disease block or death.

The same four models (demographic, comorbidity, embedding, and combined) described above were fitted for each outcome. The same MLP architecture and training procedure were used for the morbidity-accumulation endpoints. Model discrimination was evaluated using held-out test-set AUC.

For these endpoints, uncertainty in paired AUC differences was estimated by nonparametric bootstrap resampling of held-out test-set predictions.

For visualization, predicted event probabilities were converted to predicted event-free probabilities, where higher values indicate lower predicted risk of morbidity accumulation or death. 

As a sensitivity analysis for the composite endpoint, we refitted the embedding and comorbidity models after adding days since last pre-landmark admission to both feature sets.

\subsection*{Embedding Residual-Risk Analysis}

To examine what information the learned embedding contributed beyond established comorbidity burden, we performed a residual-risk analysis for the time-to-second-new-disease-block-or-death endpoint. For each individual in the held-out test set, we computed the predicted event risk from the comorbidity model and from the embedding model. The comorbidity model included age, sex, and Elixhauser score, whereas the embedding model included age, sex, and the learned patient-history embedding. Embedding residual risk was defined as the difference between these two predicted risks:
\[
r_i = \hat p_i^{\mathrm{embedding}} - \hat p_i^{\mathrm{comorbidity}}.
\]
Positive values indicate that the embedding model assigned higher risk than expected from age, sex, and Elixhauser score alone.

To compare individuals with similar baseline predicted risk, residual-risk quintiles were assigned within strata defined by age band (40--49, 50--59, 60--69, and 70--79 years), sex, and deciles of comorbidity-model predicted risk. We then assessed observed event rates across residual-risk quintiles. For descriptive characterization, we compared individuals in the lowest and highest residual-risk quintiles, referred to as embedding-low and embedding-high, respectively. We summarized pre-landmark history features including prior hospital visits, distinct diagnosis codes, distinct disease blocks, ICD-10 chapters, disease blocks per year, diagnoses per visit, and days since last admission. Differences in continuous history features were reported as standardized mean differences comparing embedding-high with embedding-low individuals. We also compared the prevalence of prior ICD-10 chapters between the two groups.

For residual-risk visualization, Kaplan--Meier curves were plotted within age bands for patients in the lowest and highest residual-risk quintiles. Separation between these groups was quantified using restricted mean event-free survival time over the five-year follow-up horizon.

\subsection*{Statistical Analysis}

Model performance was assessed using held-out test-set AUC. Confidence intervals for paired AUC gains were estimated by nonparametric bootstrap resampling of held-out test-set predictions. For morbidity-accumulation endpoints, bootstrap intervals are reported for paired AUC differences. Event-free survival was visualized using Kaplan--Meier curves for the lowest and highest residual-risk quintiles within age bands, and separation between groups was summarized by restricted mean event-free survival time over the follow-up horizon (2010--2014). Descriptive residual-risk, prior-diagnosis prevalence, and principal-component analyses were performed on held-out test-set predictions unless otherwise stated.

\section*{Data availability}
Hospital claims data involve individual health information and cannot be shared. Aggregated formats can be provided upon reasonable request.
\section*{Code availability}
Code for the model is available in the repository at \url{https://github.com/Devetak/visit-transformer}
\section*{Acknowledgments}
The project was co-funded by the Federal Ministry for Innovation, Mobility and Infrastructure under project number GZ 2023-0.841.266.
\section*{Author information}
PK and KL conceptualized the project. PK, KL and MD devised the analytic methods. MD developed the visit-level transformer. KL carried out the analysis and produced the plots and graphics. KL wrote the first draft of the manuscript. PK and MD made critical comments regarding the manuscript. All authors conducted reviewing and editing of the manuscript. All authors read and approved the final manuscript.

\section*{Supplementary information}

\subsection*{Learned embeddings capture clinically interpretable population structure}

The learned embeddings showed clinically interpretable structure before any supervised outcome modeling. PC1 explained 12.1\% of embedding variance and was strongly correlated with age at landmark (Spearman $\rho$ = 0.746), diagnosis count ($\rho$ = 0.645), Elixhauser comorbidity score ($\rho$ = 0.577), and hospital visit count ($\rho$ = 0.524), but only weakly with sex ($\rho$ = 0.071). Across PC1 deciles, mean age increased from 32.8 to 89.5 years, mean Elixhauser score from 0.03 to 10.09, mean visit count from 1.81 to 9.17, diagnosis count from 2.79 to 20.39, and death during follow-up from 0.06\% to 8.1\%.

These results indicate that the contrastive model learned a dominant age, morbidity, and utilization gradient from hospital diagnosis histories alone, before seeing any downstream outcome labels. This structure was not simply a recovery of an existing burden score. The Elixhauser score provides a one-dimensional hand-crafted summary of diagnosed comorbidity burden, whereas the learned embedding retains additional axes of variation. Consistent with this, visit count and Elixhauser score were only moderately correlated ($\rho$ = 0.454) and showed partly distinct spatial gradients across embedding space, suggesting that the embedding is not reducible to either healthcare utilization or cross-sectional comorbidity burden alone.

\begin{longtable}{p{4.5cm} p{11cm}}
\caption{Definition of Elixhauser comorbidities based on ICD-10 codes using the Quan adaptation~\cite{quan_coding_2005}.}
\label{tab:elixhauser_icd10}\\
\toprule
Comorbidity & ICD-10 codes \\
\midrule
\endfirsthead

\toprule
Comorbidity & ICD-10 codes \\
\midrule
\endhead

\bottomrule
\endfoot

Congestive heart failure & I099, I110, I130, I132, I255, I420, I425, I426, I427, I428, I429, I43, I50, P290 \\

Cardiac arrhythmias & I441, I442, I443, I456, I459, I47, I48, I49, R000, R001, R008, T821, Z450, Z950 \\

Valvular disease & A520, I05, I06, I07, I08, I091, I098, I34, I35, I36, I37, I38, I39, Q230, Q231, Q232, Q233, Z952, Z953, Z954 \\

Pulmonary circulation disorders & I26, I27, I280, I288, I289 \\

Peripheral vascular disorders & I70, I71, I731, I738, I739, I771, I790, I792, K551, K558, K559, Z958, Z959 \\

Hypertension, uncomplicated & I10 \\

Hypertension, complicated & I11, I12, I13, I15 \\

Paralysis & G041, G114, G801, G802, G81, G82, G830, G831, G832, G833, G834, G839 \\

Other neurological disorders & G10, G11, G12, G13, G20, G21, G22, G254, G255, G312, G318, G319, G32, G35, G36, G37, G40, G41, G931, G934, R470, R56 \\

Chronic pulmonary disease & I278, I279, J40, J41, J42, J43, J44, J45, J46, J47, J60, J61, J62, J63, J64, J65, J66, J67, J684, J701, J703 \\

Diabetes, uncomplicated & E100, E101, E109, E110, E111, E119, E120, E121, E129, E130, E131, E139, E140, E141, E149 \\

Diabetes, complicated & E102, E103, E104, E105, E106, E107, E108, E112, E113, E114, E115, E116, E117, E118, E122, E123, E124, E125, E126, E127, E128, E132, E133, E134, E135, E136, E137, E138, E142, E143, E144, E145, E146, E147, E148 \\

Hypothyroidism & E00, E01, E02, E03, E890 \\

Renal failure & I120, I131, N18, N19, N250, Z490, Z491, Z492, Z940, Z992 \\

Liver disease & B18, I85, I864, I982, K70, K711, K713, K714, K715, K717, K72, K73, K74, K760, K762, K763, K764, K765, K766, K767, K768, K769, Z944 \\

Peptic ulcer disease & K257, K259, K267, K269, K277, K279, K287, K289 \\

AIDS/HIV & B20, B21, B22, B24 \\

Lymphoma & C81, C82, C83, C84, C85, C88, C96, C900, C902 \\

Metastatic cancer & C77, C78, C79, C80 \\

Solid tumor without metastasis & C00, C01, C02, C03, C04, C05, C06, C07, C08, C09, C10, C11, C12, C13, C14, C15, C16, C17, C18, C19, C20, C21, C22, C23, C24, C25, C26, C30, C31, C32, C33, C34, C37, C38, C39, C40, C41, C43, C45, C46, C47, C48, C49, C50, C51, C52, C53, C54, C55, C56, C57, C58, C60, C61, C62, C63, C64, C65, C66, C67, C68, C69, C70, C71, C72, C73, C74, C75, C76, C97 \\

Rheumatoid arthritis/collagen vascular diseases & L940, L941, L943, M05, M06, M08, M120, M123, M30, M310, M311, M312, M313, M32, M33, M34, M35, M45, M461, M468, M469 \\

Coagulopathy & D65, D66, D67, D68, D691, D693, D694, D695, D696 \\

Obesity & E66 \\

Weight loss & E40, E41, E42, E43, E44, E45, E46, R634, R64 \\

Fluid and electrolyte disorders & E222, E86, E87 \\

Blood loss anemia & D500 \\

Deficiency anemia & D508, D509, D51, D52, D53 \\

Alcohol abuse & F10, E52, G621, I426, K292, K700, K703, K709, T51, Z502, Z714, Z721 \\

Drug abuse & F11, F12, F13, F14, F15, F16, F18, F19, Z715, Z722 \\

Psychoses & F20, F22, F23, F24, F25, F28, F29, F302, F312, F315 \\

Depression & F204, F313, F314, F315, F32, F33, F341, F412, F432 \\

\end{longtable}

\begin{table}[H]
\caption{Prediction performance for morbidity-accumulation and death endpoints. AUCs were evaluated on held-out test data. Parentheses show 95\% bootstrap confidence intervals for paired AUC gains from resampling held-out test-set predictions.
}
\label{tab:survival_endpoint_auc}
\centering
\scriptsize
\begin{tabular}{
p{2cm}
p{1.4cm}
p{1.6cm}
p{1.8cm}
p{1.6cm}
p{1.6cm}
p{2cm}
p{2cm}
}
\toprule
Endpoint & Event rate & Demographic AUC & Comorbidity AUC & Embedding AUC & Combined AUC & AUC gain embedding--comorbidity & AUC gain combined--comorbidity \\
\midrule
Death & 4.8\% & 0.837 & 0.866 & 0.879 & 0.880 & 0.0132 (0.0118, 0.0144) & 0.0140 (0.0129, 0.0152) \\

First new block or death & 79.2\% & 0.703 & 0.703 & 0.713 & 0.714 & 0.0102 (0.0092, 0.0111) & 0.0106 (0.0097, 0.0115) \\

Second new block or death & 49.4\% & 0.720 & 0.722 & 0.726 & 0.727 & 0.0037 (0.0031, 0.0042) & 0.0048 (0.0042, 0.0053) \\
\bottomrule
\end{tabular}
\end{table}

\bibliographystyle{unsrtnat}
\bibliography{references}

\end{document}